\documentclass[11pt]{article}
\usepackage{moriond,epsfig}

\def\be{\begin{equation}}
\def\ee{\end{equation}}
\def\bea{\begin{eqnarray}}
\def\eea{\end{eqnarray}}

\newcommand{\ep}{\epsilon}

\begin{document}
\vspace*{4cm}
\title{PRIMORDIAL GRAVITATIONAL WAVES AND THE LOCAL \\ B-MODE POLARIZATION OF THE CMB}

\author{JUAN GARCIA-BELLIDO}

\address{Instituto de F\'{\i}sica Te\'orica CSIC-UAM and Departamento de F\'isica Te\'orica, \\
Universidad Aut\'onoma de Madrid, Cantoblanco 28049 Madrid, Spain\\
D\'epartement de Physique Th\'eorique, Universit\'e de Gen\`eve, \\
24 quai Ernest Ansermet, CH--1211 Gen\`eve 4, Switzerland}

\maketitle\abstracts{
A stochastic background of primordial gravitational waves could be detected soon
in the polarization of the CMB and/or with laser interferometers. There are at least three
GWB coming from inflation: those produced during inflation and associated with the 
stretching of space-time modes; those produced at the violent stage of preheating
after inflation; and those associated with the self-ordering of Goldstone modes if inflation
ends via a global symmetry breaking scenario, like in hybrid inflation. Each GW background 
has its own characteristic spectrum with specific features. We discuss the prospects for
detecting each GWB and distinguishing between them with a very sensitive probe, the
local B-mode of CMB polarization. }


Cosmological Inflation~\cite{LindeBook,MukhanovBook} naturally generates a spectrum of density fluctuations
responsible for large scale structure formation which is consistent with the observed CMB 
anisotropies.\cite{Komatsu2010} It also generates a spectrum of gravitational waves, 
whose amplitude is directly related to the energy
scale during inflation and which induces a distinct B-mode polarization pattern in the 
CMB.\cite{DurrerBook} Moreover, Inflation typically ends in a violent process at preheating,\cite{preheating} 
where large density waves collide at relativistic speeds generating a stochastic background 
of GW~\cite{GWpreh} with a non-thermal spectrum characterized by a prominent peak at GHz 
frequencies for GUT-scale models of inflation (or at mHz-kHz for low scale models of inflation), 
and an amplitude proportional to the square of the mass scale driving/ending inflation. Such a 
background could be detected with future GW observatories like Adv-LIGO~\cite{ligo}, 
LISA~\cite{lisa}, BBO~\cite{bbo}, etc.
Furthermore, if inflation ended with a global phase transition, like in certain scenarios of
hybrid inflation, then there is also a GWB due to the continuous self-ordering of the Goldstone 
modes at the scale of the horizon,\cite{Krauss} which is also scale-invariant on subhorizon 
scales,\cite{FFDGB} with an amplitude proportional to the quartic power of the symmetry breaking 
scale, that could be detectable with laser interferometers as well as indirectly with the B-mode 
polarization of the CMB.\cite{GBDFFK}

\begin{figure}
\hspace{-0.6cm}
\psfig{figure=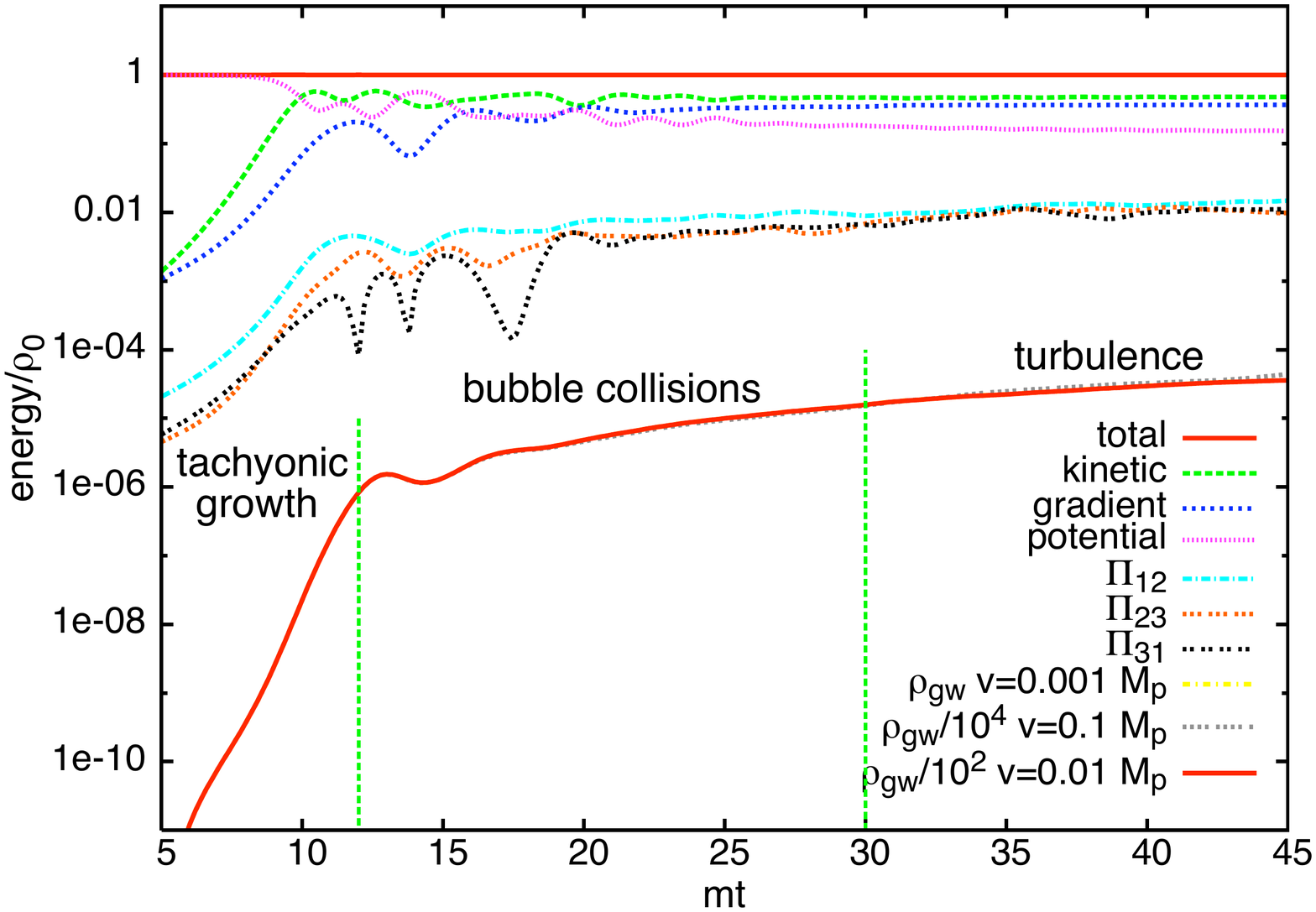,height=6.5cm}
\hspace{-1.0cm}
\psfig{figure=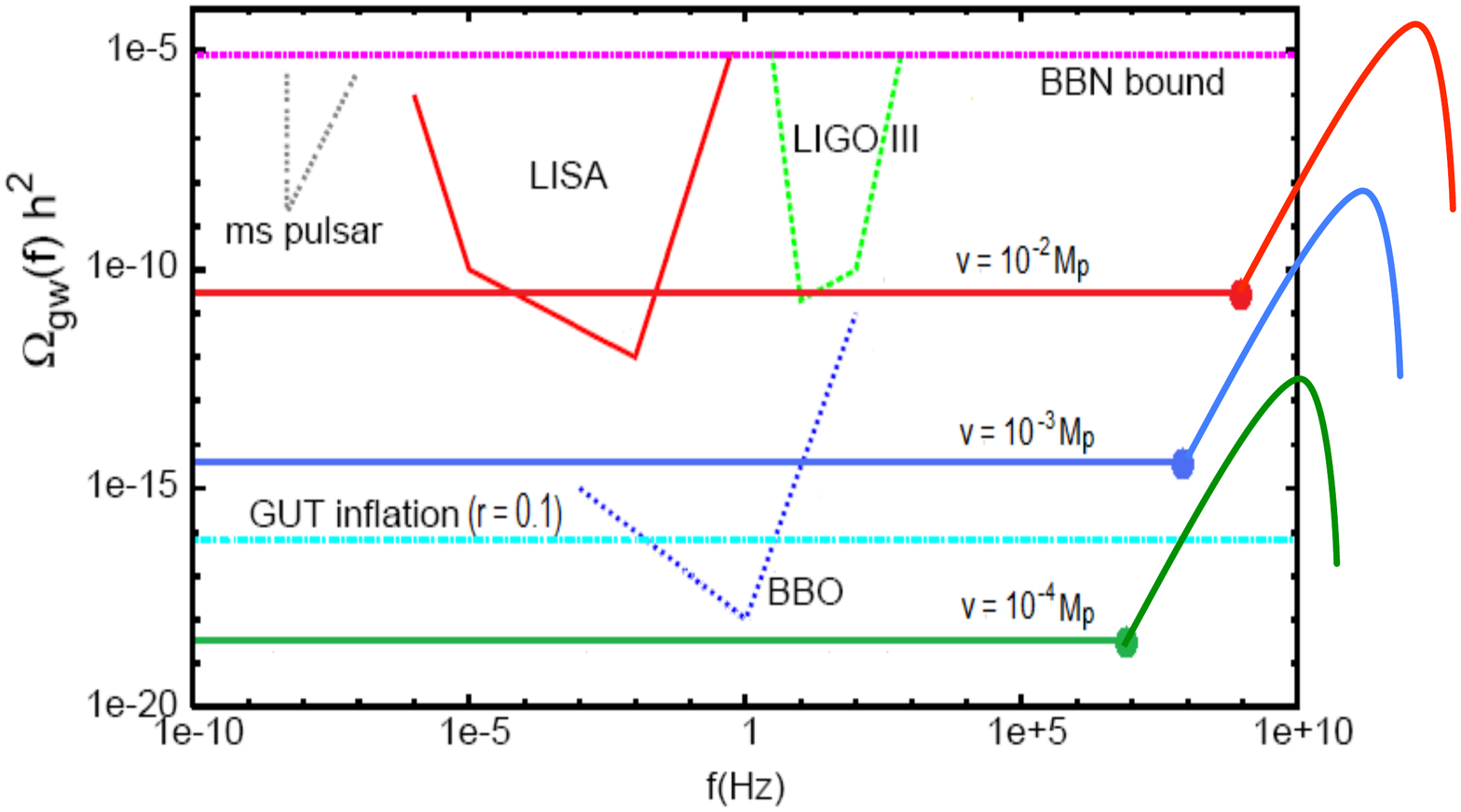,height=8.0cm,width=10cm}
\caption{LEFT: The time evolution of the GW energy density during the initial stages of preheating after
hybrid inflation, from Ref.[6]. Note the three stages of tachyonic growth, bubble collisions and turbulence.
RIGHT: The observational bounds on GW. Flat spectra correspond to GUT Inflation (dashed line) and 
global phase transitions (continuous lines),
while the peaks at the end of the latter spectra correspond to preheating at high scales.
\label{fig:GWpreh}}
\end{figure}

Gravitational waves produced during inflation arise exclusively due to the quasi-exponential expansion of the 
Universe~\cite{MukhanovBook}, and are not sourced by the inflaton fluctuations, to first order in 
perturbation theory. They have an approximately scale invariant and Gaussian spectrum whose 
amplitude is proportional to the energy density during inflation. GUT scale inflation has good chances 
to be discovered (or ruled out) by the next generation of CMB anisotropies probes, Planck~\cite{Planck}
and CMBpol~\cite{CMBpol}, see Fig.~1 for present bounds.
At the end of inflation, reheating typically takes place in several stages. There is first a rapid (explosive)
conversion of energy from the inflaton condensate to the fields that couple to it. This epoch is known as 
preheating~\cite{preheating} and occurs in most models of inflation. It can be particularly violent in the
context of hybrid inflation, where the end of inflation is associated with a symmetry breaking scenario,
with a huge range of possibilities, from GUT scale physics down to the Electroweak scale. Gravitational 
waves are copiously produced at preheating from the violent collisions of high density waves moving
and colliding at relativistic speeds~\cite{GWpreh}, see Fig.~1. The GW spectrum is highly peaked at the 
mass scale corresponding to the symmetry breaking field, which could be very different from the Hubble scale.

\begin{figure}
\psfig{figure=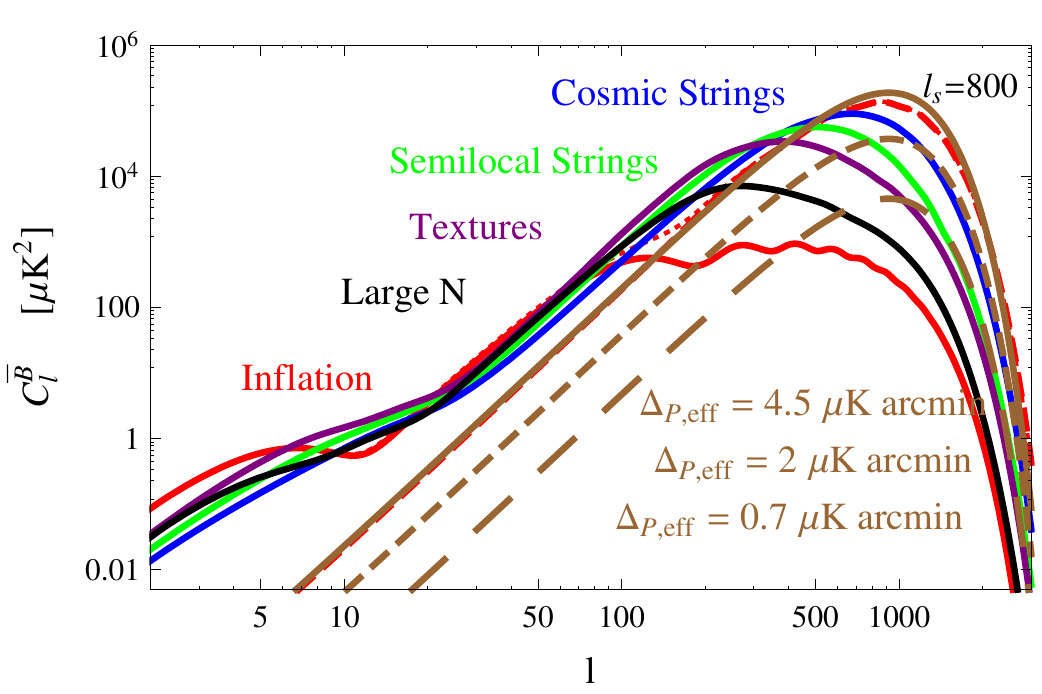,height=5.1cm}
\hspace{0.2cm}\psfig{figure=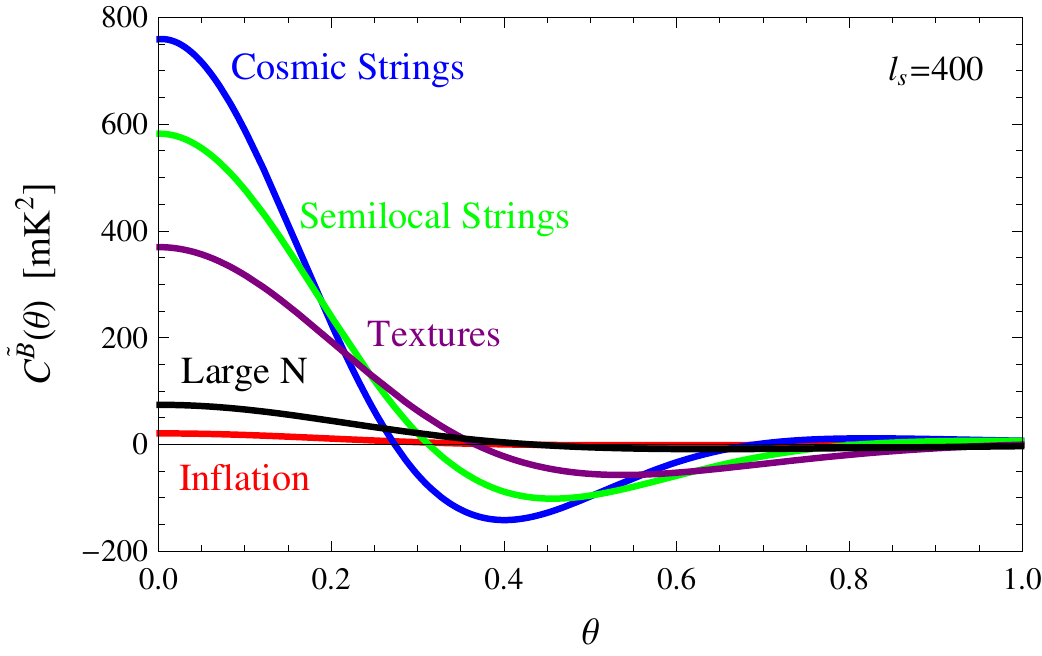,height=4.9cm}
\caption{LEFT: The local $\tilde{B}$-polarization power spectra for tensor perturbations from inflation, 
cosmic strings, textures and the large-$N$ limit of thenon-linear sigma-model. All spectra are normalized 
such that they make up 10\% of the observed temperature anisotropy at $\ell=10$. Note that the lensed EE 
modes (red dashed line) contribute as colored noise to the local-BB angular power spectrum.
RIGHT: The local $\tilde{B}$-polarization angular correlation functions for $\theta<1^o$ for inflation, 
cosmic strings, textures and the large-$N$ limit of NL sigma model.
\label{fig:CB}}
\end{figure}

In low-scale models of hybrid inflation it is possible to attain a significant GWB at the range of frequencies
and sensitivities of LIGO or BBO. The origin of these GW is very different from that of inflation. Here the
space-time is essentially static, but there are very large inhomogeneities in the symmetry breaking (Higgs) field
due to the random spinodal growth during preheating. Although the transition is not first order, ``bubbles" 
form due to the oscillations of the Higgs field around its minimum. The subsequent collisions of the 
quasi-bubble walls produce a rapid growth of the GW amplitude due to large field gradients, which 
source the anisotropic stress-tensor~\cite{GWpreh}. The relevant degrees of freedom are those of the 
Higgs field, for which there are exact analytical solutions in the spinodal growth stage, which later can be 
input into lattice numerical simulations in order to follow the highly non-linear and out-of-equilibirum stage 
of bubble collisions and turbulence. However, the process of GW production at preheating lasts only a 
short period of time around symmetry breaking. Soon the amplitude of GW saturates during the turbulent 
stage and then can be directly extrapolated to the present with the usual cosmic redshift scaling. Such a 
GWB spectrum from preheating would have a characteristic bump, worth searching for with GW observatories 
based on laser interferometry, although the scales would be too small for leaving any indirect signature in 
the CMB polarization anisotropies. Moreover, the mechanism generating GW at preheating is also active 
in models where the SB scenario is a local one, with gauge fields present in the plasma, and possibly 
related to the production of magnetic field flux tubes~\cite{PMFpreh}. In such a case, one could try to 
correlate the GWB amplitude and the magnitude and correlation length of the primordial magnetic field seed.

In the case that inflation ends with a global or local symmetry breaking mechanism, then there generically
appear cosmic defects associated with the topology of the vacuum manifold. For instance, global cosmic
strings are copiously produced during preheating if the Higgs field is a complex scalar with a U(1) global
symmetry~\cite{preheating}. In principle, all kinds of topological and non-topological defects could form at
the end of inflation and during preheating. Such defects will have contributions to all three different metric
perturbations: scalar, vector and tensor, with similar amplitudes~\cite{defrev}. 
In a recent work~\cite{FFDGB}, we analyzed 
the production of GW at preheating for a model with O(N) symmetry. The dynamics at subhorizon scales 
was identical to that of the usual tachyonic preheating. However, in this model even though the Higgs 
potential fixed the radial component to its vev, there remained the free (massless) Goldstone modes to
orient themselves in an uncorrelated way on scales larger than the causal horizon. In the subsequent 
evolution of the Universe, as the horizon grows, spatial gradients at the horizon will tend to reorder
these Goldstone modes in the field direction of the subhorizon domain. This self-ordering of the fields
induces an anisotropic stress-tensor which sources GW production. In the limit of large N components, 
it is possible to compute exactly the scaling solutions, and thus the amplitude and shape of the GWB
spectrum. It turns out that the GWB has a scale-invariant spectrum on subhorizon scales~\cite{Krauss} 
and a $k^3$ infrared tail on large scales~\cite{FFDGB}, which can be used to distinguish between inflation 
and these non-topological defects.

\begin{figure}
\psfig{figure=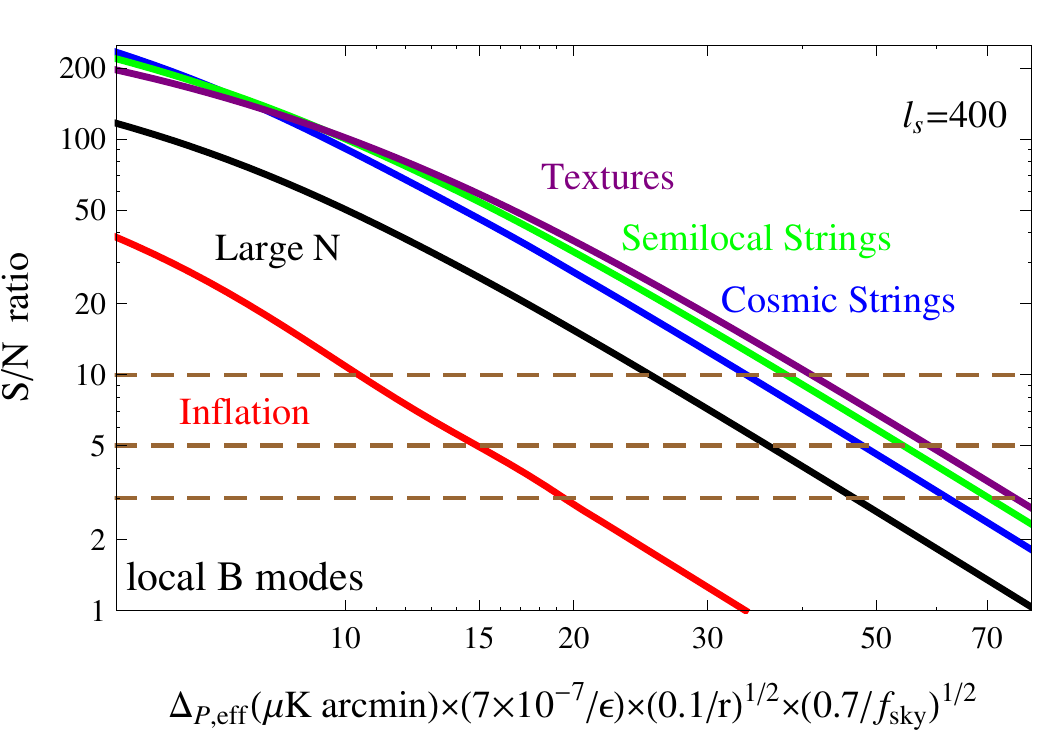,height=5.3cm}
\hspace{0.5cm}\psfig{figure=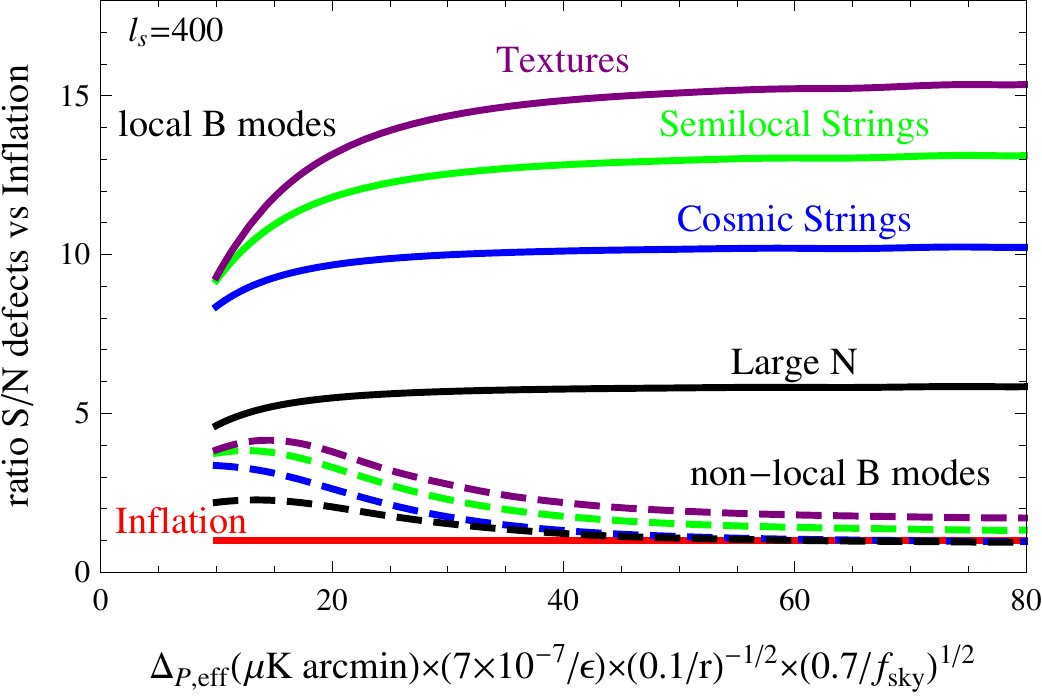,height=5cm}
\caption{LEFT: The signal-to-noise ratio as a function of the normalized polarization sensitivity,
for inflation, cosmic strings, textures and the large-$N$ limit of the non-linear sigma-model.
RIGHT: The relative signal-to-noise ratio for defect models versus inflation for local (continuous lines)
and non-local (dashed lines) B-modes.
\label{fig:SNR}}
\end{figure}

Apart from the IR tail, the main difference between inflationary and global defect contributions to the
CMB anisotropies arises from the fact that defects generically contribute with all modes: scalar, vector 
and tensor modes, with similar amplitudes, while inflationary tensor modes could be negligible if the
scale of inflation is well below the GUT scale. Since (curl) B-modes of the polarization anisotropies
only get contributions from the vector and tensor modes, the detection of the B-mode from inflation
may be challenging, and dedicated experiments like Planck and CMBpol have been designed to look 
for them. On the other hand, defects' contribution to the temperature anisotropies have a characteristic
smooth hump in the angular power spectrum, which allows one to bound their amplitude (and thus the 
scale of symmetry breaking) below $10^{16}$ GeV.~\cite{defCMB} However, the contribution to the 
B-mode coming from defects have both tensor and vector components, and the latter can be up to ten 
times larger than the former, and actually peaks at a scale somewhat below the horizon at last scattering 
(in harmonic space the corresponding multipole is $\ell\sim1000$).

In a recent paper~\cite{GBDFFK} we analyzed the possibility of disentangling the different contributions 
to the B-mode polarization coming from defects versus that from inflation. The main difficulty, for both
defects and tensor modes from inflation, is that the B-mode power spectrum is ``contaminated" by the
effect of lensing from the intervening matter distribution on the dominant E-mode contribution on
similar angular scales. Using the temperature power spectrum to determine the underlying matter
perturbation from evolved large scale structures responsible for CMB lensing, it is possible to engineer
an iterative scheme to clean the primordial B-modes from lensed E-modes~\cite{SeljakHirata}. This procedure
leaves a significantly smaller polarization noise background $\Delta_{P,{\rm eff}}$ which allows one to 
detect the GW background at high confidence level (3-$\sigma$) if the scale of inflation or that of
symmetry breaking is high enough. What we realized is that the usual E- and B-modes used for
computing the angular power spectra are complicated non-local functions of the Stokes parameters,
involving both partial differentiation and inverse laplacian integration. Such a non-local function
requires knowledge of the global polarization on scales as large as the horizon, where the B-mode
angular correlation function is negligible and thus prone to large systematic errors. In contrast, the
so-called ``local" \~E- and \~B-modes~\cite{DurrerBook,BZ} can be constructed directly from the Stokes 
parameters and do not involve any non-local inversion. A direct consequence in this change of
variables is that the angular power spectrum of local \~B-modes has a extra factor 
$n_\ell = (\ell+2)!/(\ell-1)! \sim \ell^4$, which boosts the high-$\ell$ peak in the defects' power spectra. 
When compared with the angular correlation function of inflation, it gives a significant advantage to the
defects' prospects for detection in future CMB experiments, see Fig.~3 and Table~1.


\begin{table}[h]
\caption{The limiting amplitude for inflation (r=T/S) and various defects ($\ep=Gv^2$), 
at 3-$\sigma$ in the range $\theta\in[0,1^o]$, for Planck ($\Delta_{P,{\rm eff}}=
11.2\,\mu$K$\cdot$arcmin), CMBpol-like exp. ($\Delta_{P,{\rm eff}}=0.7\,\mu$K$\cdot$arcmin) 
and a dedicated CMB experiment with $\Delta_{P,{\rm eff}}=0.01\,\mu$K$\cdot$arcmin. We 
have assumed $f_{\rm sky}=0.7$ for all CMB experiments. \label{tab:3sigma}}
\vspace{0.4cm}
\begin{center}
\begin{tabular}{|c|ccccc|}
\hline
$S/N=3$ & Inflation & Strings & Semilocal & Textures & Large-N \\
\hline
\hline
Planck & $0.03$ & $1.2\cdot10^{-7}$ & $1.1\cdot10^{-7}$ & $1.0\cdot10^{-7}$ & $1.6\cdot10^{-7}$  \\
\hline
CMBpol & $10^{-4}$ & $7.7\cdot10^{-9}$ & $6.9\cdot10^{-9}$ & $6.3\cdot10^{-9}$ & $1.0\cdot10^{-8}$  \\
\hline
$\tilde B$ exp & $10^{-7}$ & $1.1\cdot10^{-10}$ & $1.0\cdot10^{-10}$ & $0.9\cdot10^{-10}$ & $1.4\cdot10^{-10}$  \\
\hline
\end{tabular}
\end{center}
\end{table}

\section*{Acknowledgments} I thank Ruth Durrer, Martin Kunz, Elisa Fenu and Daniel Figueroa
for a very enjoyable collaboration.
This work is supported by the Spanish MICINN 
under project AYA2009-13936-C06-06.

\end{document}